\documentclass[preprint, twocolumn]{aastex63}
\usepackage{rotating}
\usepackage{amssymb}
\usepackage{amsmath}
\usepackage{color}
\usepackage{lineno}

\submitjournal{ApJ}
\shorttitle{Absorption Toward Three Local Group Dwarfs}
\shortauthors{Qu \& Bregman}

\begin{document}

\title{Absorption Line Search Through Three Local Group Dwarf Galaxy Halos}


\correspondingauthor{Joel N. Bregman}
\email{jbregman@umich.edu}

\author[0000-0002-2941-646X]{Zhijie Qu}
\affiliation{Department of Astronomy \\
University of Michigan \\
Ann Arbor, MI  48109  USA}

\author[0000-0001-6276-9526]{Joel N. Bregman}
\affiliation{Department of Astronomy \\
University of Michigan \\
Ann Arbor, MI  48109  USA}

\begin{abstract}
Dwarf galaxies are missing nearly all of their baryons and metals from the stellar disk, presumed to be in a bound halo or expelled beyond the virial radius.  The virial temperature for galaxies with $M_{\rm h} \sim 10^9 - 10^{10}$ $M_{\odot}$ is similar to the collisional ionization equilibrium temperature for the \ion{C}{4} ion.  We searched for UV absorption from \ion{C}{4} in six sightlines toward three dwarf galaxies in the anti-M31 direction and at the periphery of the Local Group ($D \approx$ 1.3 Mpc; Sextans A, Sextans B, and NGC 3109). The \ion{C}{4} doublet is detected in only one of six sightlines, toward Sextans A, with $\log N(\mbox{\ion{C}{4}})$ = $13.06 \pm 0.08$.  This is consistent with our gaseous halo models, where the halo gas mass is determined by the cooling rate, feedback, and the star formation rate; the inclusion of photoionization is an essential ingredient.  This model can also reproduce the higher detection rate of \ion{O}{6} absorption in other dwarf samples (beyond the Local Group), and with \ion{C}{4} only detectable within $\sim 0.5R_{\rm vir}$.  \\
\end{abstract}

\section{Introduction}

Several lines of studies reveal that galaxies have lost most of their baryons and most of their metals.
The ``missing baryons" problem is the finding that the mass of stars and cold disk gas is 
only about one quarter of the cosmic baryon mass expected for the dynamical mass of a galaxy \citep{mcgaugh10}.  The fraction of baryons declines toward lower mass galaxies, where dwarfs may only account for only $5\%$ of their baryons.  Closely related is the “missing metals” problem, where about $70-80\%$ of the metals formed by galaxies are present in the stars and cold gas \citep{Peep14}. 

A likely solution is that the baryons and metals exist as a gaseous halo surrounding the galaxy, possibly extending beyond the virial radius.  This requires strong feedback due to supernovae and active galactic nucleus (AGN), which is predicted to have a profound effect on the gaseous halo (e.g., \citealt{opp18b, Davies2020}.  Many searches focus on the more massive galaxies ($M_{\rm h}>10^{10} M_\odot$), which should have more extended halos, and there have been a number of successes (e.g., \citealt{Stocke2013,werk14}).  In massive galaxies, the ultraviolet (UV) absorption lines occur at temperatures below virial, so the gas that they trace is not the volume filling component.

The high ionization UV lines can trace the volume-filling medium in lower mass galaxies, the targets of this program. 
The \ion{C}{4} ion can produce strong UV absorption lines and typically is found in gas at $6 \times 10^4$ K in collisional ionization equilibrium, which corresponds to the virial temperature for a $M_{\rm halo} \sim 10^9$ $M_{\odot}$ system.  Such systems correspond to dwarf galaxies that are less massive than the Small Magellan Cloud (SMC) but more massive than the very faint dwarf galaxies in the Local Group.  It might be possible to detect such hot halos or winds around dwarf galaxies using ions like \ion{C}{4} or \ion{Si}{4}.  In this study, we search for evidence of these absorption lines around three dwarf galaxies in the Local Group, NGC 3109, Sextans A, and Sextans B (Fig. \ref{fig:GlongGlat3dwarfs}).  These galaxies were chosen because they are far from the Milky Way and M31 ($D = 1.33$ Mpc), and they are in the anti-M31 direction.  In this direction, the velocity difference between the Sun and these galaxies is $300-400$ km s$^{-1}$, largely due to the infall of the Milky Way toward M31.  This provides enough velocity separation that one can hope to identify absorption as being due to the dwarf halos.  

\subsection{The NGC 3109 Association of Dwarfs}

NGC 3109 is the most massive of the galaxies in this “group”, with $M_* + M(\mbox{\ion{H}{1}}) = 6.7 \times 10^8 M_{\odot}$, or about $10^{-2}$ of the equivalent Milky Way value, although with the important difference that the \ion{H}{1} mass is significantly larger than the stellar mass \citep{Car2013}, as $M_* \approx 8 \times 10^7\, M_{\odot}$. 
It is classified as a Magellanic-type spiral that is close to being edge-on, and with a rotation curve that is flattening by 8.25 kpc (their last data point), with an implied dynamical mass of at least $8.6 \times 10^{9}\, M_{\odot}$.  The \ion{H}{1} gas is more extended than the stellar disk, which is typical of a field spiral, and it has ongoing star formation at a rate of $\sim 5 \times 10^{-3}\, M_{\odot}\, \rm yr^{-1}$.  

The dwarf irregular galaxies Sextans A and Sextans B are similar in that they have the same luminosity and stellar mass ($M_* \approx 4 \times 10^7\, M_{\odot}$) and dynamical mass ($1 \times 10^9\, M_{\odot}$; \citealt{Bellaz2014}).  In Sextans A, the \ion{H}{1} gas is less extended than the asymmetric stellar light distribution and has $M(\mbox{\ion{H}{1}}) = 6.2 \times 10^7\, M_{\odot}$, which is slightly larger than the stellar mass.  In Sextans B, the \ion{H}{1} and stellar light are nearly identical in size and shape, where $M(\mbox{\ion{H}{1}}) = 4.1 \times 10^7\, M_{\odot}$, about equal to the stellar mass.  Star formation is active in both systems, with similar rates during the past Gyr of  $\sim2 \times 10^{-3}\, M_{\odot}\,\rm yr^{-1}$ \citep{Weisz2011}. 

\begin{figure}
  \begin{center}
	\includegraphics[width=0.49\textwidth]{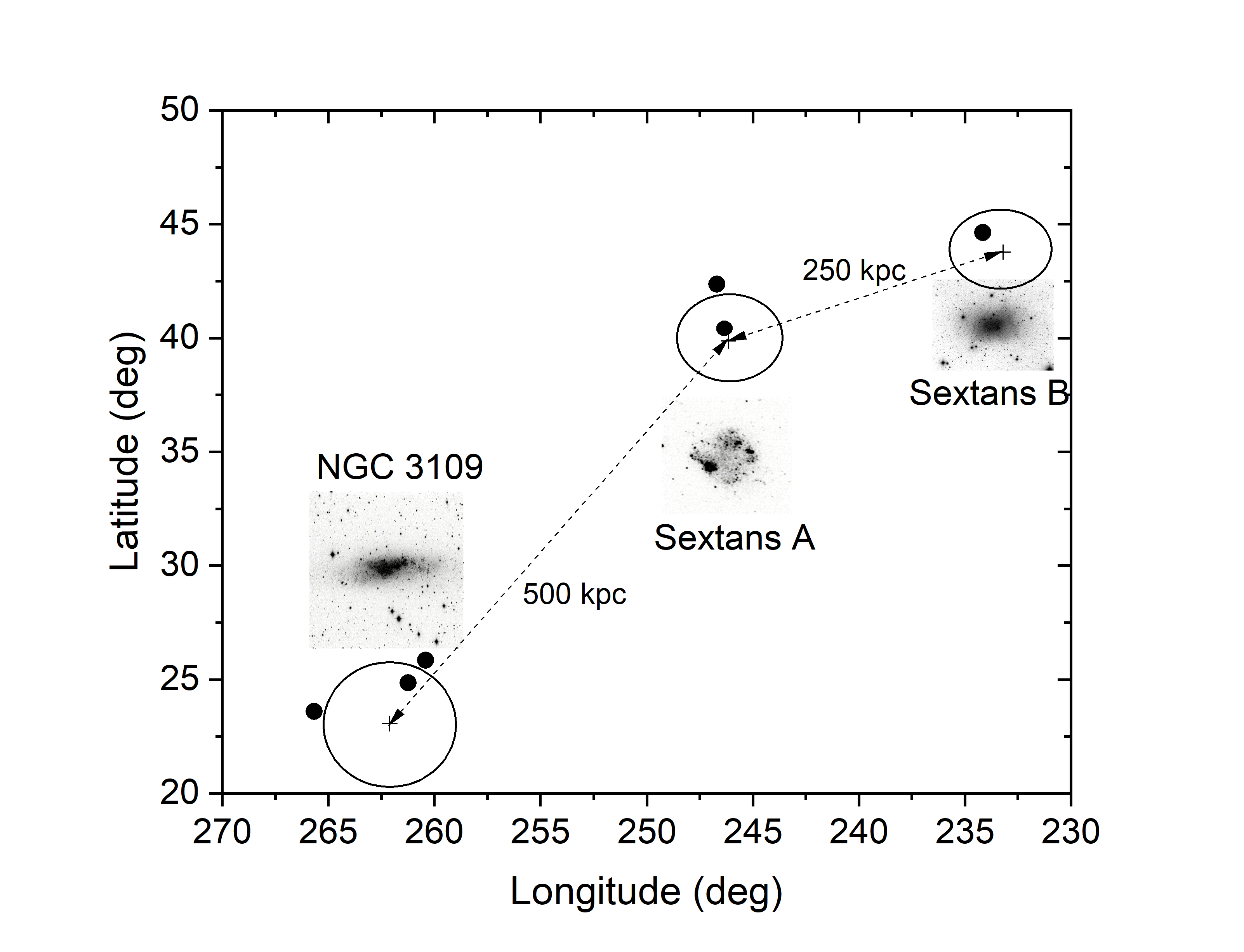}	
  \end{center}	
\vskip -0.2in
\caption{ 
The locations of the three dwarfs in this program in Galactic coordinates, with their virial radii (solid flattened circles), the five AGN sightlines (black points), and the projected separation between galaxies (dashed lines).
}
\label{fig:GlongGlat3dwarfs}
\end{figure}

NGC 3109, Sextans A, and Sextans B are at nearly the same distance and part of the sky, which might be coincidental, but \cite{Tully2006} showed that many, if not most of the dwarfs in the Local Group lie in associations, and sometimes in well-defined planes \citep{Libe2015}.  The association containing NGC 3109, Sextans A, and Sextans B has at least two other confirmed and five other possible nearby members \citep{Tully2006, Bellaz2013}. These galaxies are separated by distances ($250-500$ kpc) that are large compared to their virial radii ($40-80$ kpc; Fig. \ref{fig:GlongGlat3dwarfs}), so we can view them as each being primary galaxies.

This group of galaxies may have passed closer to the center of the Local Group in the past, but without a strong interaction with an ambient medium, as ram pressure stripping has not made these galaxies \ion{H}{1}-poor \citep{Putman2021}.  The orbital period is comparable to the age of the Universe, although the dynamical models are not in perfect agreement with the velocity and position of these dwarfs \citep{Banik2017,Peebles2017}.  This may be due to the gravitational influence of the Virgo Cluster \citep{Banik2017}, as these galaxies are nearly in the Hubble flow. 
Regardless of these details, these galaxies have spent more than a Gyr in a relatively poor environment, not near any massive galaxy.  A galactic wind or hot halo could have established itself in this period of relative isolation.  

\begin{deluxetable*}{lllccllllllc}
\tablecaption{Target Dwarf Galaxies and Background AGNs \label{tab:dwarf_AGNs}}
\tablewidth{0pt}
\tablehead{
\colhead{Dwarf}  & \colhead{Glong}     & \colhead{Glat}      & \colhead{$D$}     & \colhead{$v$} & \colhead{$M_K$}  & \colhead{AGN Name}              & \colhead{Glong}     & \colhead{Glat}      & \colhead{$z$} & \colhead{FUV}   & \colhead{Offset} \\
              & \colhead{deg.} & \colhead{deg.} & \colhead{Mpc} & \colhead{km s$^{-1}$}  & \colhead{mag} &                       & \colhead{deg.} & \colhead{deg.} &          & \colhead{mag}      & \colhead{kpc}      
}
\decimalcolnumbers
\startdata	
NGC 3109      & 262.1029  & 23.0706   & 1.33  & 403     & -16.3 & CTS M00.02            & 261.23032 & 24.85116  & 0.154    & 18.20  & 44     \\
              &           &           &       &         &       & ESO 499-G 041        & 260.41529 & 25.84116  & 0.012    & 18.05 & 73     \\
              &           &           &       &         &       & 1RXS J1015-2748 & 265.65866 & 23.59321  & 0.11     & 18.03 & 75    \\
Sextans A     & 246.1482  & 39.8755   & 1.32  & 324     & -15.6 & MARK 1253             & 246.69698 & 42.3647   & 0.049    & 17.23 & 64    \\
              &           &           &       &         &       & PG 1011-040 & 246.501105 & 40.748819 & 0.0583  & 16.24 &  21 \\
Sextans B     & 233.2001  & 43.7838   & 1.36  & 301     & -15.6 & PG 1001+054           & 234.16096 & 44.62218  & 0.16     & 17.11 & 28   \\
\enddata
Columns: (1) dwarf galaxy; (2) and (3) Galactic coordinates of the galaxy; (4) distance to the galaxy; (5) projected heliospheric velocity; (6) Absolute $K$ band magnitude; (7) background AGN; (8) and (9) Galactic coordinates of the AGN; (10) redshift of the AGN; (11) FUV magnitude of the AGN; (12) projected distance between the AGN and the galaxy.
\vspace{-0.5cm}
\end{deluxetable*}

\begin{table*}
\begin{center}
\caption{Observations of {\it HST}/COS UV spectra}
\label{table:obs_info}
\begin{tabular}{cccccccccc}
\hline
\hline
Target & Dwarf & RA & DEC & Exp$^a$ & $S/N^a$ & PI program  & obs. ID. & Date \\
 &  & h m s & d m s & ks & & &  &  \\
 (1) & (2) & (3) & (4) & (5) & (6) & (7) & (8) & (9) \\
\hline
CTS M00.02 & NGC3109 & 10 05 32.7 & $-24$ 17 16 & 5.5 7.8 & 8.0 5.8 & Bregman 13347 & LCCV030 & 2014 Apr 13\\
1RXS J1015-2748 & NGC3109 &  10 15 59.2 & $-27$ 48 29 & 5.6 7.9 & 7.2 5.3 & Bregman 13347 & LCCV050 & 2014 Nov 23 \\
ESO 499-G 041 & NGC3109 & 10 05 55.4 & $-23$ 03 25 & 5.5 7.1 & 6.1 4.7 & Bregman 13347 & LCCV040 & 2014 Jun 05\\
MRK1253 & SextansA & 10 19 32.9 & $-03$ 20 14 & 3.2 3.8 & 10.2 8.2 & Bregman 13347 & LCCV010 & 2014 Jun 08 \\
PG 1011-040 & SextansA & 10 14 20.7 & $-04$ 18 40 & 5.3 4.7 & 27.7 15.7 & Green 11524 & LB4Q040 & 2010 Mar 26 \\
PG 1001+054 & SextansB & 10 04 20.1 & $+05$ 13 00 & 3.2 3.8 & 15.1 8.1 & Bregman 13347 & LCCV020 & 2014 Jun 18\\
\hline
\hline
\end{tabular}
\end{center}
$^a$  G130M and G160M.\\
Columns: (1) AGN name; (2) dwarf galaxy; (3) and (4) equatorial coordinates of the AGN; (5) exposure time; (6) signal-to-noise ratio; (7) principle investigator and {\it HST} program; (8) observation ID; (9) observation date.
\end{table*}

\section{Data Reduction and Results}

In this study, there are six QSO sight lines around three dwarf galaxies (NGC 3109, Sextans A, and Sextans B) within the local group.
These observations are obtained by two HST programs (13347 and 11524), which is summarized in Table \ref{table:obs_info}.

The data reduction follows \citet{Wakker:2015aa} and \citet{Qu:2019aa}, and we briefly summarize the reduction steps here.
First, we co-add the gross counts obtained from individual exposures, which are corrected for the background and the fixed pattern noise.
Then, the net count rates are calculated from the gross counts by dividing the co-added effective exposure times. 
The co-added noise is the Poisson noise derived from the total counts, which is found to be better matched with the measured noise of the co-added spectrum than the co-added noise \citep{Wakker:2015aa}.
The final coadded spectrum is in the heliospheric frame.
The $S/N$ ratios of the co-added spectra are also summarized in Table \ref{table:obs_info}.

There is a known instrumental shift issue for the wavelength solution calculated from CALCOS for the COS spectrum \citep{Wakker:2015aa}.
Here, we performed a wavelength calibration for the co-added spectra, which employs the \ion{H}{1} 21 cm lines as references for UV absorption lines.
The \ion{H}{1} 21 cm emission lines are extracted from the Leiden/Argentine/Bonn (LAB) Survey of Galactic \ion{H}{1} with an effective beam size of $0.2^\circ$ around the QSO sight lines \citep{Kalberla:2005aa}.
The Galactic absorption lines of low ionization state ions (e.g., \ion{Si}{2} and \ion{C}{2}) are assumed to have the same velocity line centroid as the \ion{H}{1} 21 cm lines, and a set of velocity shifts are obtained over the entire spectrum.
Then, we fit polynomial functions to the obtained velocity shifts simultaneously for the four spectrum segments in both G130M and G160M.
In most cases, we only use a constant or linear functions to correct the wavelength, and only the segment B of G130M of PG 1001+054 use a second order polynomial.
The co-added spectrum is binned by three pixels for fitting and line identification, yielding the bin width of $\Delta \lambda =0.02991~\rm \AA$ and $0.03669~\rm \AA$ for G130M and G160M, respectively.

\begin{figure*}[ht]
  \begin{center}
	\includegraphics[width=15cm]{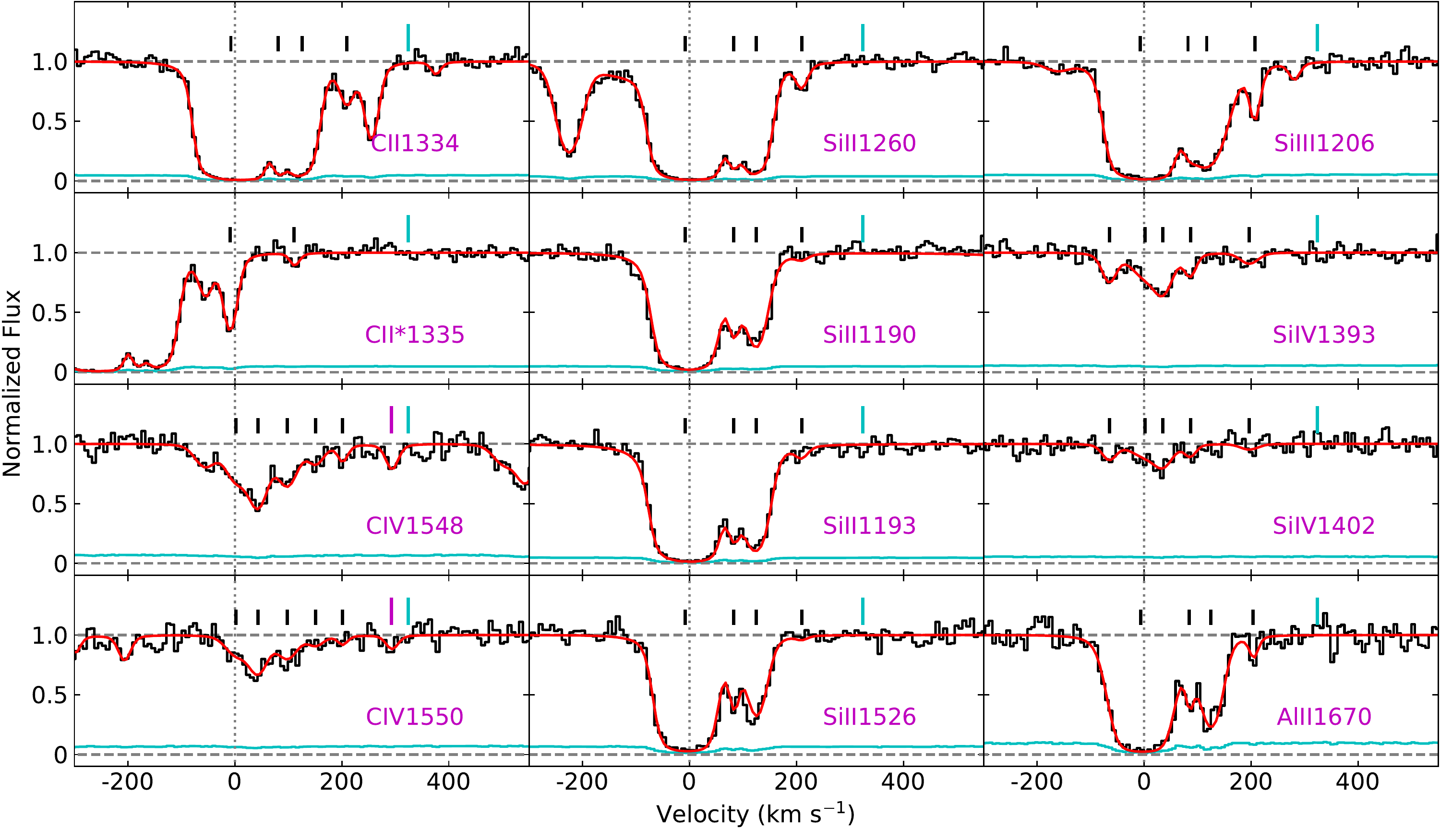}	
  \end{center}	
\vskip -0.2in
\caption{
The \textit{HST}/COS spectra for the strongest lines from common ions along the sight line of PG 1011 around Sextans A, showing the sum of all model components (red), the data (black), and the observation uncertainty (cyan).  Individual absorption components are marked above the continuum.  Unmarked features are from the contamination of other lines (e.g., the red-most line in the \ion{C}{2} panel at the top-left).  One finds a weak \ion{C}{4} doublet at 294 km s$^{-1}$ (marked in a long magenta bar), 30 km s$^{-1}$ from the systemic velocity of Sextans A (324 km s$^{-1}$ in long cyan bars). 
}
\label{fig:SexAspectra}
\end{figure*}

Most lines are non-detection.
The equivalent widths (EWs) and uncertainties are calculated over a typical width of $60~\rm km~s^{-1}$, which is about three times the typical $b$ values ($20~\rm km~s^{-1}$) measured for cool-warm CGM.
The only detection is \ion{C}{4} of PG 1011-040 for the Sextans A at heliocentric $v=294 \rm~ km~s^{-1}$ (Fig. \ref{fig:SexAspectra}), and $-30\rm~ km~s^{-1}$ from the systemic velocity of Sextans A.
We fit the Voigt profile for this \ion{C}{4} detection, obtaining the column density of $\log N = 13.04 \pm 0.08$ and $b=9.9\pm5.3\rm~km~s^{-1}$.
The 2 $\sigma$ upper limits of the column densities, are calculated from EW uncertainties by assuming $b=20\rm~km~^{-1}$ for other none-detection (Table \ref{table:column}).

In this sight line of PG 1011-040, there is also a possible \ion{Si}{3} $\lambda 1206.5\rm~\AA$ feature at $v_{\rm helio} = 280 \rm~km~s^{-1}$, which is different in velocity from the $294 \rm~km~s^{-1}$ \ion{C}{4} system by $3\sigma$.
We do not consider it as an absorption feature associated with Sextans A for two reasons.
First, \ion{Si}{3} only has one transition at $1206.5\rm~\AA$ in the {\it HST}/COS coverage, and no other ions show absorption features at a similar velocity.
Thus, it may be contamination from intervening absorption systems beyond the Local Group.
Second, this feature is $44\rm~km~s^{-1}$ away from the bulk velocity from Sextans A, which is beyond the rotation velocity of the galaxy ($\approx 30 \rm~km~s^{-1}$; \citealt{Namumba2018}).

\begin{table*}
\caption{Column Measurements of Ions}
\begin{center}
\label{table:column}
\begin{tabular}{ccccccccc}
\hline
\hline
Galaxy & AGN & $\rho$ & \ion{C}{2} & \ion{C}{4} & \ion{O}{1} & \ion{Si}{2} & \ion{Si}{3} & \ion{Si}{4}  \\
(1) & (2) & (3) & (4) & (5) & (6) & (7) & (8) & (9) \\
\hline
NGC 3109 & CTS M00.02 & 44 & $13.6$ & $13.3$ & $...^a$ & $12.7$ & $12.7$ & $12.6$\\
& ESO 499-G 041 & 73 & $13.6$ & $13.5$ & $...^a$  & $12.6$& $12.7$ & $13.0$\\
& 1RXS J1015-2748 & 75 & $13.6$ & $13.3$ & $14.4$ & $12.5$ & $12.5$ & $13.0$\\
Sextans A & MRK 1253 & 64 & $...^a$ & $13.3$ & $...^a$ & $12.2$ & $12.4$ & $12.8$\\
& PG 1011-040 & 21 & $...^a$ & $13.04\pm 0.08$ & $13.4$ & $11.8$ & $11.9$ & $12.3$\\
Sextans B & PG 1001+054 & 28 & $...^a$ & $13.1$ & $13.8$ & $12.2$ & $12.1$ & $12.6$\\
\hline
\hline\\
\end{tabular}
\end{center}
$^a$ Affected by geocoronal emission or Galactic absorption.\\
Columns: (1) dwarf galaxy; (2) AGN; (3) impact parameter in kpc; (4) - (9) log of the column densities of ions, where the column is in units of cm$^{-2}$.
\end{table*}

\section{Discussion and Conclusions}

Our only detection is in the sight line closest to Sextans A, where the separation is 55$^{\prime}$, or 21 kpc, well within the virial radius, $\approx 44$ kpc (Fig. \ref{fig:GlongGlat3dwarfs}).  This can be compared to the outermost radius at which the \ion{H}{1} emission is detected in LITTLE THINGS (the Very Large Array, VLA) and by the Square Kilometre Array (SKA) pathfinder KAT-7 \citep{Hunter2012, Namumba2018}.  The two observations yield the same radial profile and have a limiting 3$\sigma$ \ion{H}{1} column density for the VLA observation is $2 \times 10^{18}$ cm$^{-2}$ and a value of $5.8 \times 10^{18}$ cm$^{-2}$ for the KAT-7 observation. The \ion{H}{1} is detected to a radius of 3.5 kpc (9$^{\prime}$), which is six time smaller than the sight line-galaxy separation.  This indicates that the absorption is more likely due to the gaseous halo around Sextans A rather than an extension of the disk. 

\subsection{Comparison to Other Dwarf Surveys}
 
There are four studies of dwarf galaxies beyond the Local Group that can be used for comparison.  They differ somewhat in selection criteria and redshift space, and for $M_*$ similar to our galaxies, are either at $z < 0.02$ \citep{Stocke2013,Bord2014, Burchett2016} or $z \approx 0.2$ \citep{Johnson2017}.  The \ion{C}{4} absorption detection rate is similar between these surveys, and for $7.5 < \log M_* < 9$, the detection rate is 3/37 galaxies, summed over the four surveys.  For the detected systems, the impact parameter is $< 0.5R_{\rm vir}$, and a similar result is found for more massive dwarfs, $9 < \log M_* < 10$ \cite{Burchett2016}.

We acknowledge differences between the sample but find that the results of these surveys are consistent with the absorption results in this work, where 1/6 sightlines have a \ion{C}{4} detection, and that sight line is the closest to the galaxy (Sextans A).  The probability of having one detection for six sightlines is about $50\%$, assuming a Poisson probability of 3/37 per sightline, so this our detection rate is in agreement with the surveys. 

It is worth noting that the absorption line detection rate is larger for \ion{H}{1} and \ion{O}{6} (ions that we cannot observe in our sample).  \cite{Johnson2017} finds that \ion{H}{1} is detected in nearly every galaxy (17/18) and that \ion{O}{6} is detected in one-third of the galaxies (6/18).  We return to this result for \ion{O}{6} when discussing absorption models for dwarf galaxies (below). 

\begin{figure}
  \begin{center}
	\includegraphics[width=0.48\textwidth]{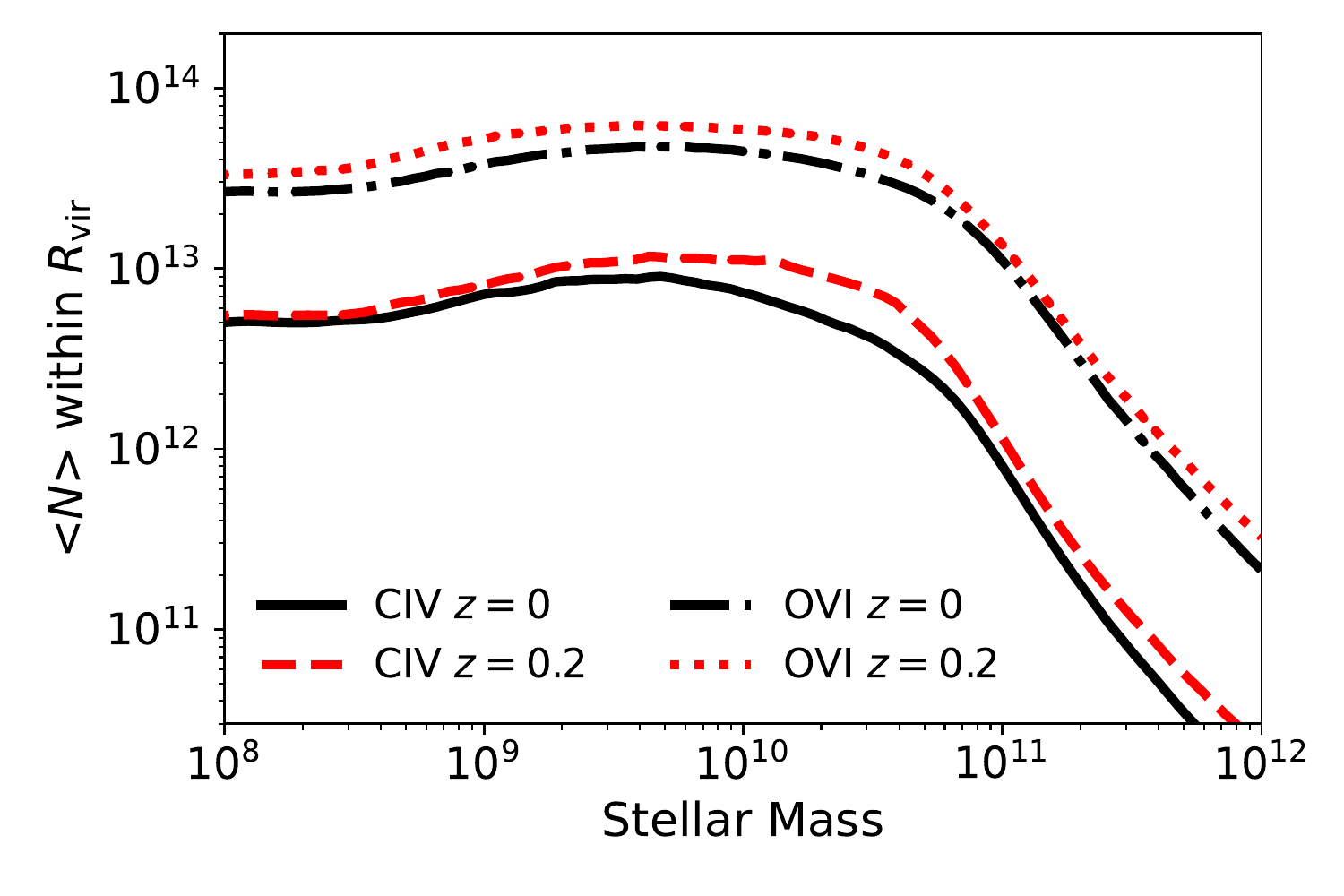}	
  \end{center}	
\vskip -0.2in
\caption{The mean \ion{C}{4} and \ion{O}{6} column densities within $R_{vir}$ as a function of stellar mass in the TPIE model.  The effect of photoionization leads to the extended flat part of the distribution to lower $M_*$.
}
\label{fig:NCIVOVImodel}
\end{figure}

One can also consider some of the observations toward other dwarfs in the Local Group.  One of the problems with such studies, including ours, is that Galactic gas comes in various forms (e.g., high-velocity cloud, intermediate-velocity cloud, and Magellanic Stream), which covers a range of velocities \citep{Richter2017}.  This can lead to a confusion problem that one tries to resolve by using velocity separation from the known gaseous features in the halo.  

The dwarfs IC 1613 and WLM are in the same part of the sky as the Magellanic Stream and within $30-40^{\circ}$ of M33 and M31.  In WLM, \cite{Zheng2019WLM} find absorption components that include \ion{C}{4} at velocities of $-150$ and $-220$ km s$^{-1}$.  These velocities can be compared to the systemic velocity of WLM ($-132$ km s$^{-1}$), and the Magellanic Stream, where the mean velocity is $<-190$ km s$^{-1}$.  They tentatively assign the $-150$ km s$^{-1}$ system to the halo of WLM.

\begin{figure*}
  \begin{center}
	\includegraphics[width=8cm]{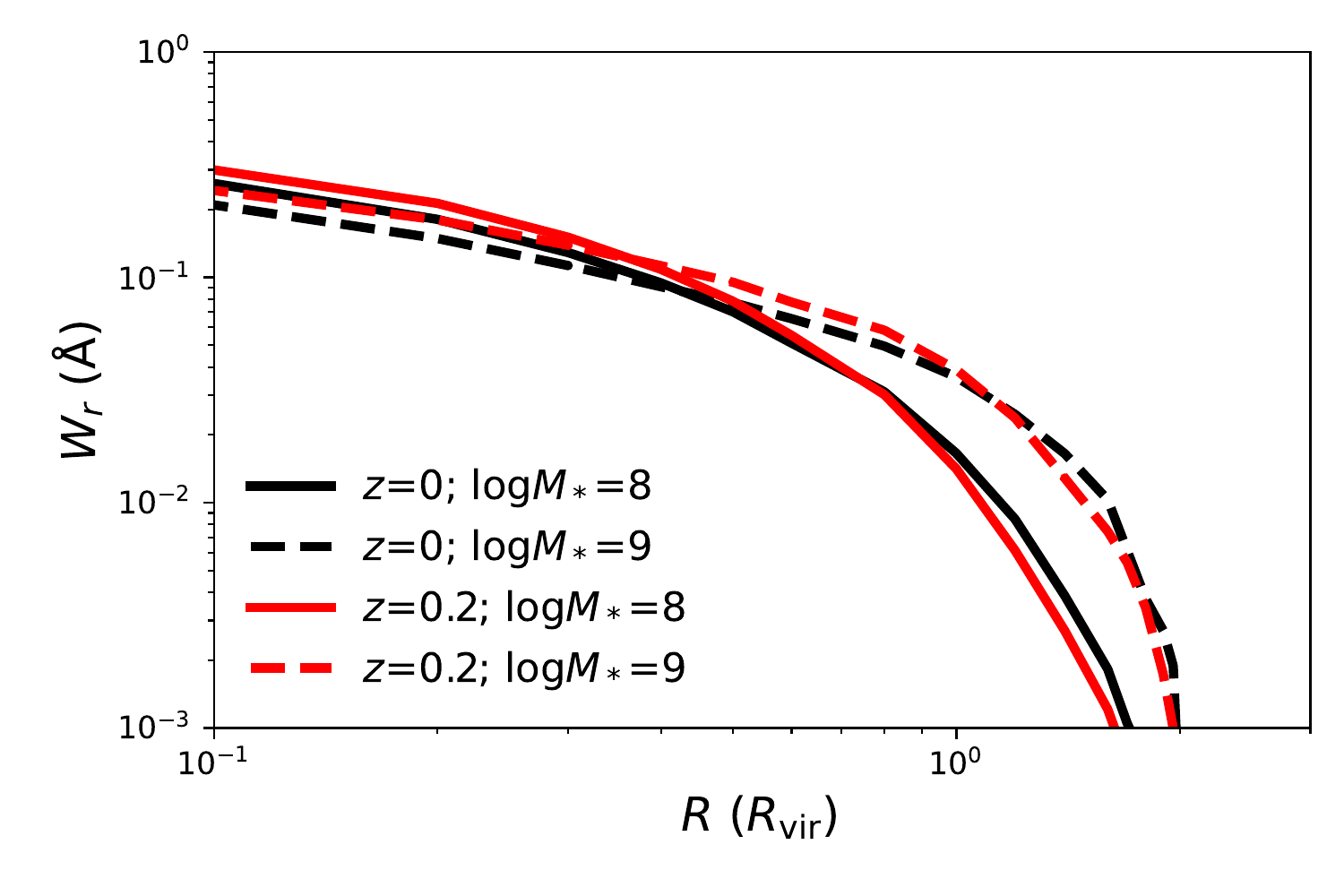}	
	\includegraphics[width=8cm]{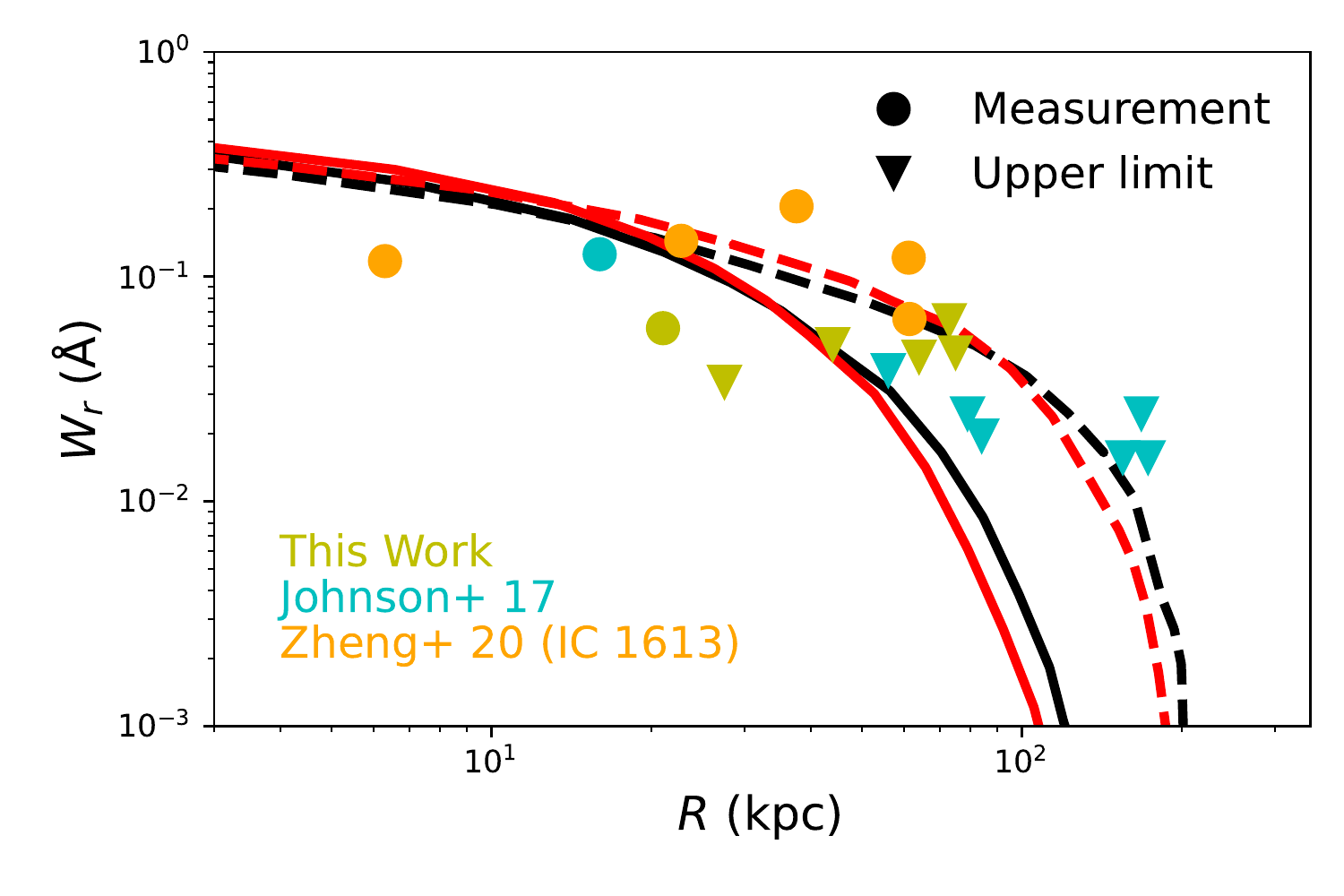}	
  \end{center}	
\vskip -0.2in
\caption{{\it Left panel}: the equivalent width of \ion{C}{4} $\lambda 1548$ (in \AA) versus the fractional virial radius for galaxies with two stellar masses and at two redshifts. The two distributions by stellar mass (at fixed redshift) are similar, with a sharp decline beyond $\approx0.8R_{\rm vir}$ at $z = 0$ and $\approx1.2R_{\rm vir}$ at $z = 0.2$.  The difference is mainly due to the increased star formation rate at higher redshift. 
{\it Right panel}: the equivalent width of \ion{C}{4} $\lambda 1548$ (in \AA) vs the projected radius in kpc for the three $z = 0$ galaxies in this study, which have log$M_*/M_\odot = 7.6-8.6$.  The only detection is for Sextans A (log$M_*/M_\odot = 7.6$ at the smallest radius; downward triangles are upper limits. The points for \cite{Johnson2017} are in the range log$M_*/M_\odot = 7.7-9.2$ and at $z \approx 0.2$.  The model seems to be consistent with the data for log$M_*/M_\odot \approx 8$  but might over-predict the columns for log$M_*/M_\odot \approx 9$.
}
\label{fig:WCIV}
\end{figure*}

The dwarf IC 1613 has four sightlines toward hot stars in the dwarf and six more from AGN sight lines \citep{Zheng2020IC1613}.  This is more massive than WML, having $M_* \approx 1 \times 10^8 M_{\odot}$, $M(\mbox{\ion{H}{1}}) = 0.65 \times 10^8 M_{\odot}$, and $M_{\rm h} \approx 4 \times 10^{10} M_{\odot}$, and at a velocity shift of $-232$ km s$^{-1}$.  \cite{Zheng2020IC1613} identify absorption, typically within 100 km s$^{-1}$ of the systemic velocity of IC 1613 for all ten sightlines.  This 100\% success rate makes IC 1613 quite unusual relative to other dwarfs.  They discuss whether some of these features might be associated with the Magellanic Stream, but conclude that IC 1613 is the most likely absorption line host. 

We also consider the contamination from unassociated Galactic gas for Sextans A.  From the survey of 270 sightlines, \cite{Richter2017} shows that sightlines in this general direction ($200^{\circ} < l < 300^{\circ}$, $30^{\circ} < b < 60^{\circ}$ have absorption at velocities of $100-200$ km s$^{-1}$, which we find in our sightlines as well.  However, it is impossible to be certain that a small amount of halo gas, unrelated to Sextans A, is responsible for the absorption at 294 km s$^{-1}$.

\subsection{Comparison to Gaseous Halo Models}

One can try to estimate whether a gaseous halo around a dwarf galaxy would have enough column density in metals to be detectable.  The upper limit in the column is $N_{200} = n_{200}R_{200}$, where $n_{200}$ is 200 times the critical baryon density of the universe at $z = 0$.  For our dwarfs, this yields $N_{200} = 5-10 \times 10^{18}$ cm$^{-2}$, and for gas at 0.1 Solar metallicity, and an ionization fraction of 0.2 (near the peak), the \ion{C}{4} column density would be  $N(\mbox{\ion{C}{4}})_{200} = 3-5 \times 10^{13}$ cm$^{-2}$, which is somewhat larger than our upper limits for \ion{C}{4} of $1.3-2 \times 10^{13}$ cm$^{-2}$, or the single \ion{C}{4} detection of $1.48 \pm 0.30 \times 10^{13}$ cm$^{-2}$.  However, agreement would occur with a somewhat lower \ion{C}{4} ionization fraction, which motivates the need for more realistic models.

A more accurate and realistic estimate is possible by considering the processes likely to occur in the halos of galaxies.  This includes radiative cooling, feedback, and photoionization from the ambient metagalactic radiation field.  These processes were considered for a range of galaxy masses in \cite{Qu2018a} and \cite{Qu2018b}, which consider several models, including those with collisional ionization equilibrium and photionization equilibrium.  The most realistic model is where phototionization \citep{Haardt2012} is included in the ionic distribution and where the net cooling rate of the gaseous halo equals the star formation rate in the galaxy.  For star formation, we use a mean value for a galaxy of mass $M_{\rm h}$ and a typical feedback rate (details are given in the above papers).  A multi-temperature halo gas results from the cooling and the feedback, which we denote as the TPIE model (defined in \citealt{Qu2018a}).  Here we extend the mass of the halos to values lower than previously published, along with column densities for both \ion{O}{6} and \ion{C}{4}.

In the TCIE model (collisional ionization equilibrium in the multitemperature gas), the \ion{O}{6} column density turns sharply downward for log$M_{\rm h} < 11.4$, as the virial temperature falls below the peak \ion{O}{6} ionization fraction.   However, in the TPIE model, that turndown at lower $M_{\rm h}$ does not occur because photoionization raises the fractional ionization (Fig. \ref{fig:NCIVOVImodel}).  The result is that the \ion{C}{4} column density varies by less than a factor of two for 7.5 $<$ log$M_*/M_\odot < 10.5$ ($z < 0.2$), and rarely rises above a column of $10^{13}$ cm$^{-2}$.  This is similar for the \ion{O}{6} ion, although the column density is larger by a factor of three to six.  There is scatter in these column densities, as galaxies differ in their star formation rate and halo mass.  In general, the model suggests that the typical $N(\mbox{\ion{C}{4}})$ column density lies below the {\it HST}/COS detection limit in these surveys, while $N(\mbox{\ion{O}{6}})$ is close to the detection threshold.

We also calculated the equivalent widths of the \ion{C}{4} $\lambda 1548$ as a function of with radius (Fig. \ref{fig:WCIV}).  There is a decline in the equivalent width with radius, typically by a factor of three from $0.1-0.5R_{\rm vir}$, and another factor of two to five from $0.5-1.0R_{\rm vir}$, depending on mass and redshift (Fig. \ref{fig:WCIV}).  This helps to explain why absorption detection are rarely found beyond $\approx 0.5$ $R_{\rm vir}$, although more detectable absorption lines were expected for the dwarfs with log$M_*/M_\odot \approx 9$. 

To summarize, the TPIE models produce predictions for the column density of \ion{C}{4} and \ion{O}{6} that reproduce some of the features found in studies of absorption of dwarfs beyond the Local Group.  This suggests that dwarf galaxies contain extended gaseous halos, provided they are not in environments where their halo has been stripped by the ambient hot medium of a more massive galaxy.


\acknowledgements
We would like to thank the \textit{Hubble Space Telescope} program for their support through program HST-GO-13347.  The authors also acknowledge insights provided by Mario Mateo, Sean Johnson, Jiangtao Li, Edmund Hodges-Kluck, Mike Anderson, and a helpful anonymous referee. 

\newpage 

\bibliography{main}{}
\bibliographystyle{aasjournal}

\end{document}